\documentclass[journal12pt,draftclsnofoot,onecolumn]{IEEEtran}
\usepackage{amsmath}
\usepackage{bm}
\usepackage{amssymb}
\usepackage{latexsym}
\usepackage{multirow}
\usepackage{epsfig}
\usepackage{graphics}
\usepackage{mathrsfs}
\usepackage{algorithm}
\usepackage{balance}
\usepackage{algpseudocode}
\usepackage{subcaption}
\usepackage{color}
\usepackage{graphicx}
\usepackage{algcompatible}
\usepackage{array}
\usepackage{amsfonts}
\usepackage{mdwmath}
\usepackage{mdwtab}
\usepackage{eqparbox}
\usepackage{epstopdf}
\newcommand{\bc}{\begin{center}}
\newcommand{\ec}{\end{center}}
\newcommand{\be}{\begin{equation}}
\newcommand{\ee}{\end{equation}}
\newcommand{\bea}{\begin{eqnarray}}
\newcommand{\eea}{\end{eqnarray}}

\newcounter{tempEquationCounter}
\newcounter{thisEquationNumber}

\begin{document}

\title{Adaptive Federated Learning With Gradient Compression  in Uplink NOMA}
\author{Haijian Sun, \emph{Member, IEEE}, Xiang Ma, \emph{Student Member, IEEE}, and Rose Qingyang Hu, \emph{Fellow, IEEE} \\

\thanks{H. Sun is with the Department of Computer Science, University of Wisconsin-Whitewater, Whitewater, WI, USA. (email: h.j.sun@ieee.org).

X. Ma and R. Q. Hu are with Electrical and Computer Engineering Department, Utah State University, Logan, UT, USA. (email: xiang.ma@ieee.org,  rose.hu@usu.edu). }

%\thanks{The research was supported in part by National Science Foundation under
%grants EARS 1547312.}
}
\maketitle

\begin{abstract}
Federated learning (FL) is an emerging machine learning technique that aggregates model attributes from a large number of distributed devices. Several unique features such as energy saving and privacy preserving make FL a highly promising learning approach for power-limited and privacy sensitive devices. Although distributed computing can lower down the  information amount that needs to be uploaded,  model updates in FL can still experience performance bottleneck, especially for updates via wireless connections. In this work, we investigate the performance of FL update with mobile edge devices that are connected to the parameter server (PS) with practical wireless links, where uplink update from user to PS has very limited capacity.
Different from the existing works, we apply non-orthogonal multiple access (NOMA) together with gradient compression in the wireless uplink. Simulation results show that our proposed scheme can significantly reduce aggregation latency while achieving similar accuracy. 
 
\end{abstract}


\section{Introduction}
Machine learning (ML) techniques have achieved remarkable performance in various applications such as object detection and content recommendation, etc \cite{cache}. State-of-the-art ML exploits the growing computation power of mobile devices that are capable of collecting, sharing, and processing data. Even though such devices are still  resource-constrained compared with dedicated high performance computing (HPC) centers, crowd-sourcing a  large number of them can build powerful ML models. 
Thus, one can foresee that ML on distributed devices is one of the promising trends \cite{book}. 

Distributed ML requires devices to periodically share model attributes. Communication can become a severe challenge, especially when wirelessly connected devices participate distributed ML. Recently, federated learning (FL) can alleviate this challenge \cite{fedavg}  by selecting a small portion of the  devices and also only trained models for updating.  FL  not only  is communication efficient but also can preserve privacy and handle heterogeneous data.  In wireless systems, non-orthogonal multiple access (NOMA) is an emerging  access technique for 5G.  It allows multiple users to share the same time/frequency resource simultaneously. To minimize the impact from interference, successive interference cancellation (SIC) is applied at the receiver side, which starts the decoding for the signal with the strongest power, then subtracts the decoded signal from the composite message \cite{uplink}. The process is  sequentially carried out until the intended  signals are obtained.  

To further realize more efficient communication, in \cite{SHan}, deep gradient compression is applied to reduce the message size for transmission.  It was found that $99.9$\% of the gradient exchange is redundant. \cite{tian} proposed a modification to tackle FL in heterogeneous networks.  The  above two works did not consider the constraint of the actual communication process.  The work  presented in \cite{comm}  exploited the medium access control (MAC) property and used both analog and digital communications to directly get  the model average. However, they did not consider the effects from wireless fading channels. Besides, gradient projection transmitted over MAC directly leads to higher bit error rate (BER). In this work,  we use NOMA in the uplink communication \cite{uplink} by considering channel fading and adaptively compress gradient for the optimum transmission. The contributions are summarized as follows. 
\begin{itemize}
\item To the best of our knowledge, this is the first paper that utilizes NOMA in 5G and beyond wireless systems for FL model update. The study shows that NOMA is a favorable  selection in such a scenario compared with traditional time division multiplexing access (TDMA) approach.  

\item We consider the capacity-limited fading channel in the model and propose to apply adaptive compression techniques in the uplink NOMA.

\item We demonstrate the effectiveness of the proposed scheme with several datasets. The results show that the communication latency for FL update is reduced by at least 7x without loss of accuracy. 

\end{itemize}

This paper is organized as follows. Section II introduces the  system model, FL update mechanism, and NOMA transmission scheme. Section III presents two adaptive compression techniques for FL model aggregation. Simulation results are shown in Section IV, where we conduct extensive experiments to verify the proposed schemes. Lastly, Section V concludes and gives future research direction. 
\section{System Model}
In the system we consider a total of $N$ edge devices that distributively and  collaboratively  build a global learning model. Each device, or user, contains its own raw data locally, denoted as $\mathcal{D}_n$ for user $n$. For each dataset $j$, it can be partitioned into two parts: a vector $\mathbf{x}_j$ that represents features of data, and a vector $\mathbf{y}_j$ that is the output or label of the data. ML generally finds the mapping between $\mathbf{x}_j$ and $\mathbf{y}_j$. To analyze,  we use the function $f(\mathbf{x}_j, \mathbf{y}_j; \bm{\theta}_j)$ to capture the error between this mapping. Here $\bm{\theta}_j$ is the parameter set and $f$ is the loss function. Some typical loss functions can be linear regression and root-mean-squared error. 

Each user performs machine learning locally. Essentially, local learning aims to solve the following problem:
\be
\min_{\bm{\theta}}F_n(\bm{\theta}) = \frac{1}{|\mathcal{D}_n|}  \sum_{j \in \mathcal{D}_n}  f(\mathbf{x}_j, \mathbf{y}_j; \bm{\theta}_j), 
\ee
where $|\mathcal{D}_n|$ is the cardinality of the dataset $\mathcal{D}_n$. 
\subsection{FL Model Update}
Different from the traditional learning process where $\{ \mathcal{D}_1, \mathcal{D}_2, \ldots, \mathcal{D}_K\}$ are placed in the same location, FL relies on distributed stochastic gradient (DSG) to perform update in each iteration. Specifically, the loss function in (1) can be generalized across multiple devices  as: 
\be
\min_{\bm{\theta}}f(\bm{\theta}) =  \sum_{n=1}^N \frac{|\mathcal{D}_n |}{ \mathcal{D}}  F_n (\bm{\theta}),
\ee
where $|\mathcal{D}| = \sum_{n=1}^N |\mathcal{D}_n|$. 

For FL, in each training round, it selects a portion of the  devices to participate the  global update. Let $K = C N$ be the total number of participants, and $ 0< C < 1$.  PS initializes the model as $\bm{\theta}^0$ and sends it to all the users. Each user performs local training and calculates the gradient $\mathbf{g}_k = \nabla F_k(\bm{\theta})$. In the FL setting, however, each user can apply multiple iterations of gradient calculation. For example, in round $t$, user $k$ calculates $\bm{\theta}_k^{t} =  \bm{\theta}_k^{t} - \eta \nabla F_k(\bm{\theta})$ multiple times, where $\eta$ is the step size or the  learning rate. Participating users then send their gradients to the PS for aggregation. The PS further calculates $\bm{\theta}^{t+1} = \bm{\theta}^{t} -  \sum_{k=1}^K \frac{|\mathcal{D}_k|}{ \mathcal{D}} \bm{\theta}_k^t$  and  sends $\bm{\theta}^{t+1}$ to all the users for the next round update. In a typical wireless setting, traditional update uses TDMA for uplink transmission. The PS needs to wait until receiving the last user's message and then  averages the received information from all the users.  

\subsection{Uplink NOMA Transmission}
Unlike the traditional TDMA update, NOMA allows multiple users to share the uplink channel simultaneously. Assume that the channel between user $k$ and the PS to be $h_k$ and $h_k = L_k h_0$, where $L_k$ is the large-scale fading and $h_0$ is the small-scale fading. To simplify the analysis, we assume $L_k$ follows the free-space path loss model $L_k = \frac{\sqrt{\delta_k} \lambda }{4 \pi d_k^{\alpha/2}}$, $\delta_k$ accounts for the transmitter and receiver antenna gain, $\lambda$ is the signal wavelength, $d_k$ is the distance between user $k$ and the PS, and $\alpha$ is the path-loss exponent. Furthermore, $h_0 \sim \mathcal{CN} (0, 1)$ is the normal Gaussian variable. 

Let $s_{k}^t$ denote the encoded gradient update from user $k$ at  $t$, $s_{k}^t$ be the transformation from $\bm{\theta}_k^t$ in the local update stage. Additionally, we normalize the transmitted symbols $ || s_{k}^t  ||_2^2 = 1$.  According to NOMA principle, all the selected $K$ users share the same bandwidth  simultaneously. In particular, all the transmitted signals from multiple NOMA users are superposed \cite{noma}. The received signal at the PS at $t$ thus can be expressed as: 
\be
y^t = \sum_{k=1}^K \sqrt{p_k} h_k s_k^t + n^t, 
\ee
where $n^t \sim \mathcal{CN} (0, \sigma^2)$ is the imposed additive noise.  

SIC is carried out at the PS side. Specifically, PS decodes the strongest signal first by treating others  as interference. After successful decoding, PS  subtracts the decoded signal from the superposed signal. The process stops until the PS decodes all the participants' messages.  Without loss of generality, we assume $p_1 h_1^2 > p_2 h_2^2 > \ldots > p_K h_K^2$. Therefore, the achievable data rate for user $k$ is:
\be
R_k = \log_2 \bigg\{ 1+  \frac{p_k h_k^2 }{ \tau( \sum_{j = k+1}^K p_j h_j^2 + \sigma^2)  } \bigg\}, \forall k = 1, \ldots K-1,
\ee
where $\tau > 1$ accounts for performance degradation from finite length symbol, imperfect channel estimation, and decoding error, etc.  
User $K$ is the last decoded hence its rate is $R_K = \log_2 (1 + \frac{p_K h_K^2}{\tau \sigma^2} )$. 

\begin{figure}[ht]
	\includegraphics[width=5in]{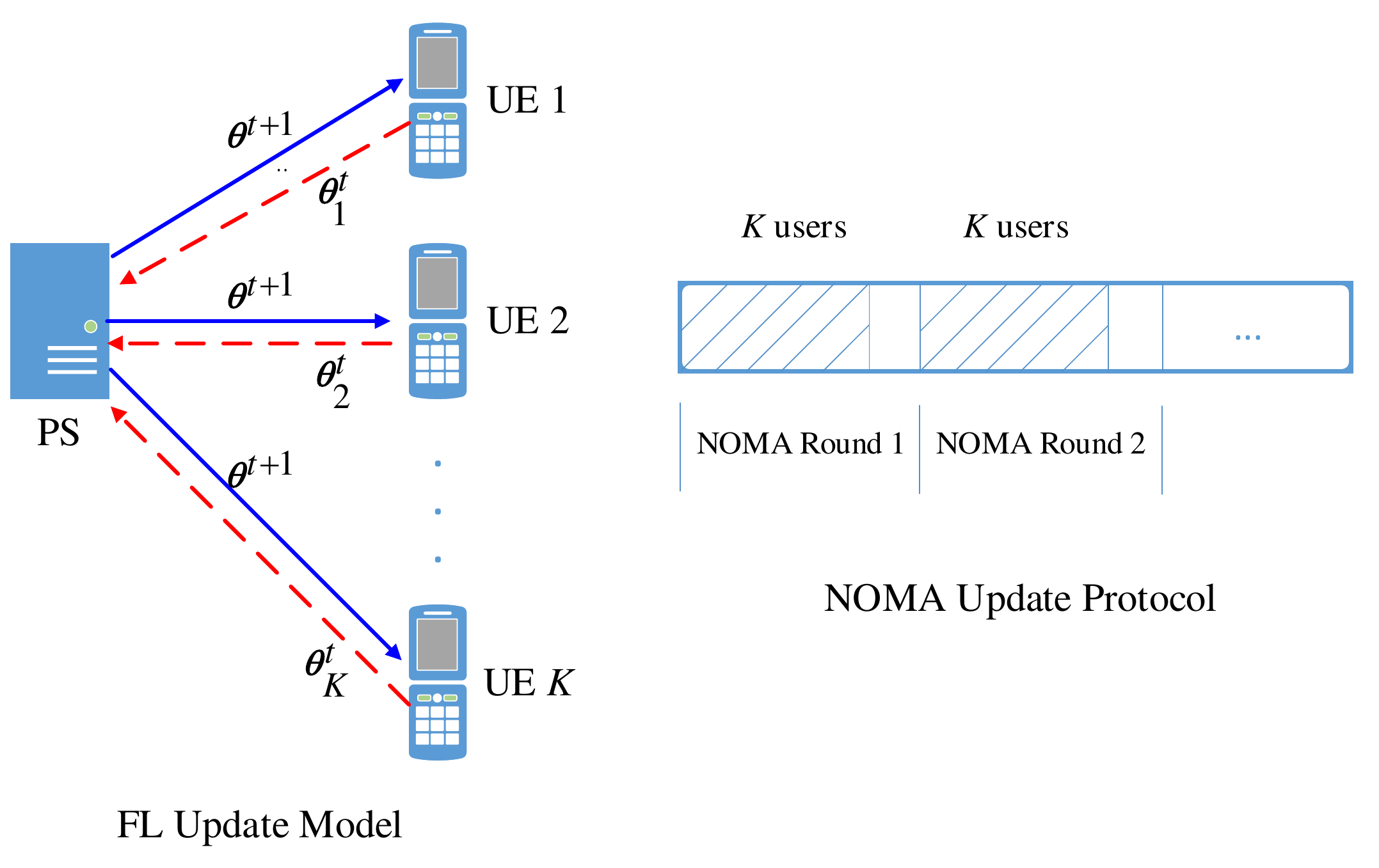}
	\centering
	\caption{ A brief illustration of the proposed scheme. Left: A general FL model update. Right: NOMA update protocol in each round. Shaded area is for uplink and blank area is for downlink.}
	\label{model}
	\centering
\end{figure}

At the beginning of each round $t$, the PS notifies the participated users to start the simultaneous transmission. The maximum number of allowable bits for user $k$ is $m_k = B R_k t_k$, where $B$ is the system bandwidth, $t_k$ is the NOMA transmission duration. A brief illustration of the FL model update and the proposed uplink NOMA is shown in Fig. \ref{model}. 

\section{Adaptive Model Compression with NOMA Transmission}
To meet the physical limitation on  the number of allowable symbols, each device should adjust their update size.  
In the case that the update size exceeds $m_k$, model compression can be applied. In this section, we briefly introduce two lossy compression algorithms and their rationales. 
\subsection{Gradient Quantization}
It is well-known that quantization can help compress the  size of a large data. Standard algorithms in machine learning typically use $32$-bit floating-point to represent each model parameter. Cost to store, transmit, and manipulate those data tends to be high. Alternatively, a simple implementation is to use less bits for such a representation. Even though quantization creates ``rounding errors'' from limited available levels, existing works have shown this approach demonstrates a good model convergence \cite{SHan}. 

In this  work, we adopt DoReFa scheme \cite{Zhou} since it is suitable for quantizing gradients within $[-1, 1]$, which is true for most ML models. Specifically, mapping can be established  with the following function
\be
q_k(x)=\frac{1}{a}\lfloor{ax}\rceil.
\ee
Here, $\lfloor\cdot\rceil$ rounds to the nearest integer, $x$ is the gradient value, and $a=2^b-1$, where $b$ is the quantization bit length.

For the selected user $k$ in each round, we calculate the maximum throughput $m_k$ under the  NOMA scheme. The total bit length of gradients $G$ is known once we have determined the ML structure, so the compression rate $r_q^k$ is calculated as $r_q^k = \max \{ \frac{G} { m_k}, 1\}$. 
The quantization bit length $b_q^k$ is calculated by $b_q^k =  \lfloor \frac{1}{r_q^k}  32  \rfloor$, 
where $\lfloor \cdot \rfloor$ takes the floor operation. Afterwards, every gradient value in user $k$ is represented by bits with a length of  $b_q^k$.
%When compression rate is high, quantization bit length tends to be small, it may cause the testing accuracy loss.

\subsection{Gradient Sparsification}
Sparsification refers to the approach that sends  selected gradients instead of sending all of them. Empirical experiments have shown that a large portion of the gradient updates in a distributed SGD are redundant. Therefore, we can first map the smallest gradients to zero and then make a sparse update. For example, the threshold-based sparsification only keeps the gradients larger than a known threshold  but sets the rest small gradients to $0$. The selection normally is based on the absolute gradient value.  After sparsification, the non-zero gradients are uploaded to the PS.

This paper employs a similar method to perform gradient sparsification. Since PS averages the gradient in an element-wise way, it is important to take the  non-zero index into consideration. Additionally, we update the relative distance (delta) between adjacent non-zero values rather than recording its absolute position.  Moreover, the well-known non-linear coding called Golomb code is applied to encode the delta value, which uses variant-length bits to further save space. The average number of bits used for encoding delta with Golomb coding is 
\be
\bar{b}_{pos}^k = b_k^{*} + \frac{1}{1-{{(1-r_s^k )^{2}}^{b_k^*}}},
\ee
where  $b^{*}=1+\lfloor \log_2(\frac{\log(\phi-1)}{\log(1-r_s^k )})\rfloor$,  $\phi=\frac{\sqrt{5}+1}{2}$, and $r_s^k$ is the spasification ratio.  $r_s^k$ can be calculated by solving the following equation
\be \label{sparse}
G  r_s^k+ \frac{G  r_s^k }{32}  \bar{b}_{pos}^k = m_{k}.
\ee
Hence $r_s^k = \min \{ r_s^k(n),1 \}$,  $ r_s^k(n)$ is the numeric solution for (\ref{sparse}). Once $r_s^k$ is obtained, we set $(1-r_s^k)$ portion of the smallest gradient values to zero.

For both compression methods above, it is important to keep the gradient residual for the next round.
  $\Delta \bm{\theta}_k^t = \bm{\theta}_k^t - GS(\bm{\theta}_k^t)$ or  $\Delta \bm{\theta}_k^t = \bm{\theta}_k^t - GQ(\bm{\theta}_k^t)$. Therefore, we can reduce the compression accumulation errors. 
 
\subsection{NOMA Scheduling}
To select $K$ users from a total of $N$ to participate the model update, we should  consider not only the learning process, but also  the communication process. Selection criteria is based on two rules: 1) NOMA fairness; 2) time budget.

1)  NOMA fairness:  From PS update, $\bm{\theta}^{t+1} = \bm{\theta}^{t} -  \sum_{k=1}^K \frac{|\mathcal{D}_k|}{ \mathcal{D}} \bm{\theta}_k^t$, the weighted average is applied, hence we use the following ``effective update capacity'' as
\be
R_{ef}^k = \frac{ B R_k t_k }{|\mathcal{D}_k|} 
\ee
to account for the actual contribution for weighted average update. As will be discussed later in Section IV, FL experiences performance degradation when data rate is heterogeneously distributed. Therefore, to make sure every user has a quality update, we use a widely accepted Jain's fairness index, which is defined as: 
\be
J_u = \frac{(\frac{1}{K} \sum_{k=1}^K R_{ef}^k)^2 }{\frac{1}{K} \sum_{k=1}^K ( R_{ef}^k)^2},
\ee
for maximum fairness, $J_u$ should be close to 1. In practice, PS  selects  users with a quality effective update capacity and ensure $J_u$ to be close to 1. We adopt a similar scheduling algorithm as in \cite{fairness}. 

2) Time budget:  For a faster update, another factor is the computation time at each device. Since NOMA is a synchronous system, at each round, we ensure each device starts the transmission simultaneously. This means PS sets a hard time budget for the local computation. Each device may have a heterogeneous capacity hence they can make the estimation for the time spent on each iteration of training. Our scheduling selects those who can not only finish the calculation on time but also make more iterations.  \textbf{Algorithm 1} summarizes the  proposed scheme. 
\begin{algorithm}[]

\caption{Adaptive FL Update with Uplink NOMA and Gradient Compression}
\begin{algorithmic}[1]
\STATE  {\bf Initialization:} PS gives initial $\bm{\theta}^0$, maximum rounds $T$. 
\FOR {each FL update round $t$} 
\STATE PS selects $K = C N$ users and calculates their maximum achievable data rates $m_k$. Then sends synchronous pilots, $m_k$, and $ \bm{\theta}^t$ to users. 
\FOR {each selected user $k$ in parallel} 
\STATE {Update local gradient one or multiple times:   $\bm{\theta}_k^{t} =  \bm{\theta}_k^{t} - \eta \nabla F_k(\bm{\theta})$, according to time budget. }
\STATE Based on $m_k$ and size of gradient, apply either sparsification $GS(\bm{\theta}_k^t)$ or quantization  $GQ (\bm{\theta}_k^t)$. 
\STATE Gradient residual $\Delta \bm{\theta}_k^t$ will be kept locally for next round update. 
\STATE send gradients to the PS at the beginning of synchronous time slot. 
\ENDFOR
\STATE PS applies SIC to decode gradient from $K$ users. 
\STATE PS performs weighted average:  $\bm{\theta}^{t+1} = \bm{\theta}^{t} -  \sum_{k=1}^K \frac{|\mathcal{D}_k|}{ \mathcal{D}} \bm{\theta}_k^t$. 
\ENDFOR

\end{algorithmic}
\end{algorithm}

\section{Numeric Evaluation} 
In this section, we present the experimental FL update results for both TDMA based original FedAvg \cite{fedavg} and NOMA compression based FedAvg schemes by using  diverse learning models and federated datasets. The channel parameters  are given as follows. We assume the bandwidth for the uplink is $B = 5$MHz, path loss exponent $\alpha = 3$, additive noise power density $\sigma^2 = -174$dBm/Hz. The number of  selected user is $K = 10$ or $K=20$ in each round. All the users are randomly distributed in a disk region of a radius $500$m and they have the same transmission power $p_k = 0.1$watts, $\forall k$.  Uplink transmission time slot is $t_k =0.5$s, $\forall k$. Downlink transmission from PS to all the users uses broadcast and is uncompressed, Transmission time is calculated as  $T_d=\max_{k} \frac{32*\mathcal{P}}{B_d \log_2 (1+P_d \gamma_k)}$, where $\mathcal{P}$ is the total parameter number, $B_d$ is the downlink bandwidth and is $10$ MHz. $P_d = 2$watts is the PS power, $\gamma_k$ is the signal-to-noise-ratio (SNR) from the PS to $k$-th user. To make the proposed scheme more convincing, we explore the convex classification problems with MNIST and Federated Extended MNIST (FEMNIST) datasets and non-convex text sentiment analysis task on tweets from Sentiment140 (\emph{Sent140}). The two image classification tasks use the same LeNet-300-100 model while \emph{Sent140} uses a long short-term memory (LSTM) classifier. Besides, under FL setting, we also employ  datasets from \cite{tian} to make the data points on different devices intentionally non-i.i.d. (number of data points per devices varies). The statistics of the datasets are summarized in Table I \cite{tian}. 
\begin{table}[h]
	\newcommand{\tabincell}[2]{\begin{tabular}{@{}#1@{}}#2\end{tabular}}
	\centering
	\caption{Statistics of Datasets}
	\begin{tabular}{c|p{19mm}|p{15mm}|p{15mm}}
		\hline
		\textbf{Dataset} & \textbf{Num. of \newline Parameters ($\mathcal{P}$)} & \textbf{Num. of \newline Devices ($N$)  } & \textbf{Num. of \newline Data ($\mathcal{D}$)  } \\
		\hline
		MNIST & \tabincell{l}{266,610 } & 1,000 & 69,035 \\
		\hline
		FEMNIST & \tabincell{l}{266,610} & 200 & 18,345\\
		\hline
		\emph{Sent140} & \tabincell{l}{243,861} & 660 & 40,783\\
		\hline
	\end{tabular}
\end{table}

For hyperparameters in FL, we use a batch size $\mathcal{B} = 10$ on all datasets. Learning rate and communication round are fixed for each dataset but may vary for different datasets. Specifically, We use a learning rate $\eta = 0.001$ and a maximum communication round $T = 100$ for MNIST,  a learning rate $\eta = 0.003$ and  a maximum communication round $T = 300$ for FEMNIST and learning rate $\eta = 0.05$ and maximum communication round $T=100$ for \emph{Sent140}. For the selected user, it uses the majority portion of data for training and the rest for testing. Testing  is performed at each device and accuracy is calculated based on the average across all the devices. 

\begin{figure}[!h]
\centering
	\includegraphics[width=90mm]{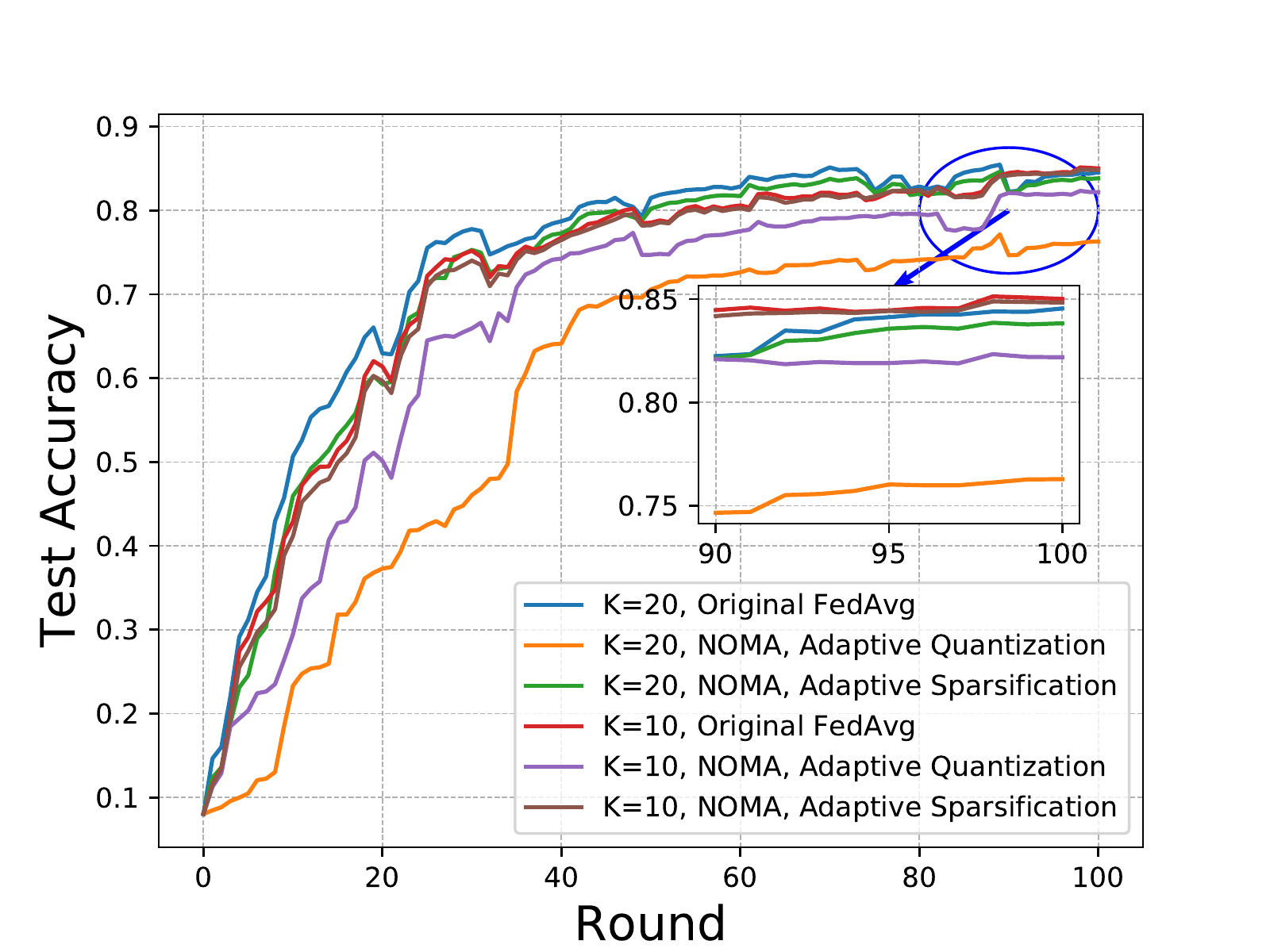}
	\caption{Test Accuracy comparison under different scenarios, when $K=10$, $K=20$, and original TDMA-based FedAvg with our proposed NOMA compression based FedAvg.}
	\label{fig:mnist_round}
	\centering
\end{figure}

In Fig. \ref{fig:mnist_round}, we present the test accuracy on MNIST dataset. As a baseline, FedAvg algorithm from \cite{fedavg} is adopted, where gradients are sent with no compression (32-bit per parameter), in a round-robin way. While in our scheme, we compress the  gradient values with either adaptive quantization or sparsification. The average compression ratio of adaptive quantization in each round is $0.55$ and $0.33$ for $K=10$ and $K=20$, respectively. For adaptive sparsification, the average compression ratio is $0.53$ and $0.31$, respectively. From Fig. \ref{fig:mnist_round}, all the schemes except quantization under $K=20$ achieves a similar accuracy (over $80$\%) at round $100$. This indicates that compression causes a minimal impact on the final results. Original FedAvg has better performance when $K=20$ since more participated users bring more overall gradient update and reduces effects from non-i.i.d. dataset. However, we observe that under NOMA transmission scheme, $K=10$ has a better result.  When $K=20$ mutual interference in NOMA causes significant data rate degradation for the first few decoded users. Thus their compression strategy becomes more aggressive, which then causes destructive effects. Lastly, comparison between quantization and sparsification is also shown. In all scenarios, sparsification has superior performance. The reason is that most of the gradient values fall into regions close to zero. When the maximum allowable data rate becomes small, quantization with low-length bit simply rounds them to $0$, which indicates no contribution for the update. 
\begin{figure}[!h]
\centering
	\includegraphics[width=90mm]{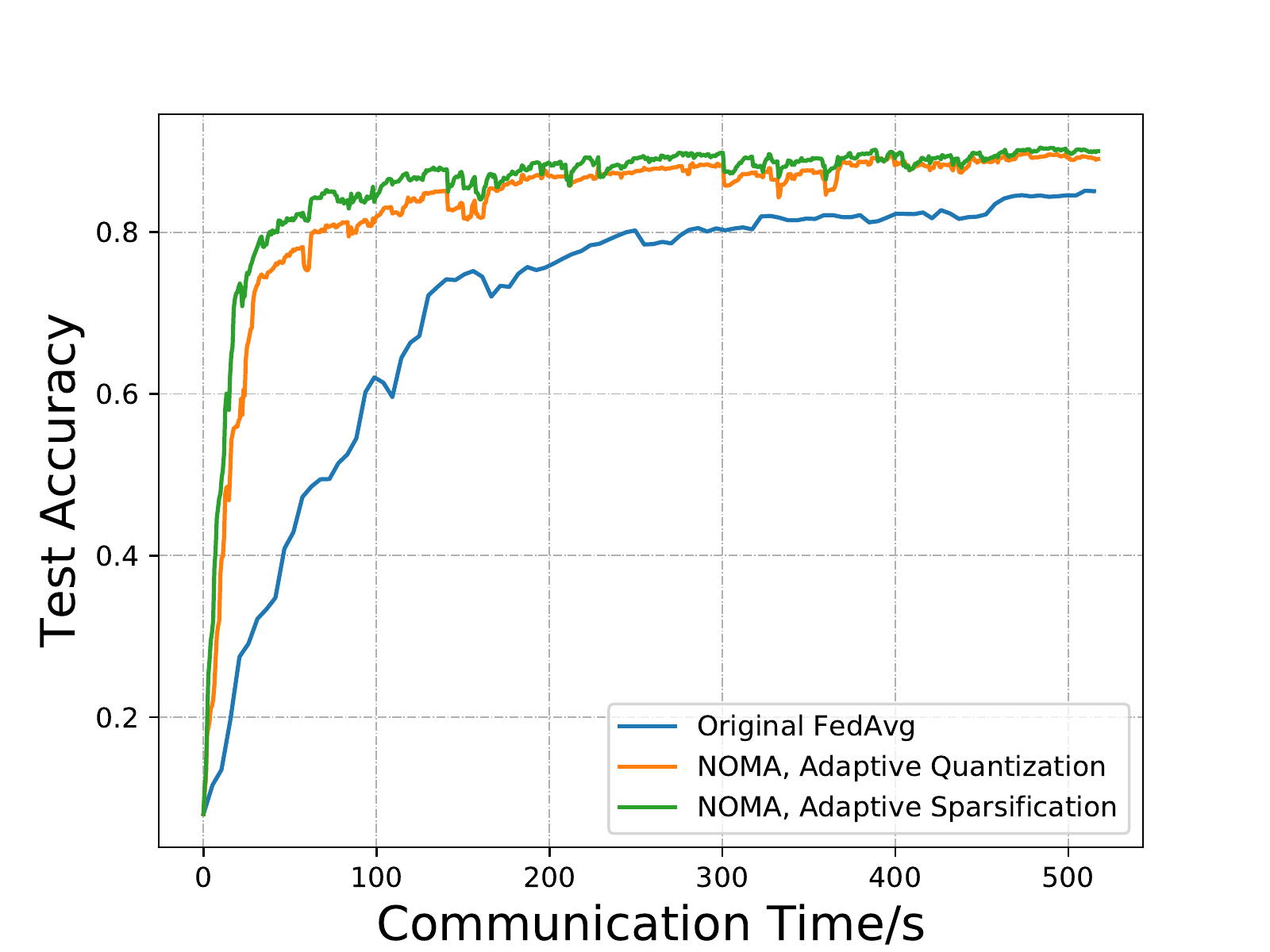}
	\caption{Test Accuracy Comparison between original TDMA-based FedAvg and NOMA compression based FedAvg update with communication time.}
	\label{fig:mnist_time}
	\centering
\end{figure}

While Fig. \ref{fig:mnist_round} shows the proposed scheme has comparable accuracy with FedAvg which takes no consideration for wireless channels. Fig. \ref{fig:mnist_time} presents a more remarkable result when comparing communication latency. For simplicity, we only show the scenario with $K=10$. In each round of NOMA compression based FedAvg update, it takes $ t_k + T_d$, while original FedAvg takes $K t_k + T_d$. It can be readily verified that, under this practical setting, NOMA compression based FedAvg update only consumes around $70$s for wireless communication to achieve 85\% of accuracy, while for the same accuracy, original TDMA-based FedAvg takes more than $500$s. NOMA-aided FL can save $7.4\times$ communication time during the update process. Alternatively, with $500$s training in NOMA protocol, we see the accuracy improves from 85\% to 88.6\% for adaptive quantization and 90\% for sparsification. Notice that under $K=20$, the time difference is more dominant. 

To generalize our proposed scheme, we run tests on FEMNIST and and \emph{Sent140}, as shown in Fig.  4 and 5, respectively. The former also uses the same LeNet-300-100 model structure as in MNIST but has different datasets and the latter uses LTSM with a non-convex loss function. Here, we only present results for $K=20$. Under all scenarios, we observe a test accuracy fluctuation, especially for FEMNIST dataset, since it is a highly non-i.i.d. scenario where each device has almost a distinct data distribution. 

Nevertheless, we can see similar performance between the  NOMA compression based FedAvg and the original FedAvg in terms of accuracy v.s. rounds, even with gradient compression. Notice that the original FedAvg and our gradient sparsification have almost the identical results in each round hence they can hardly been differentiated from figures. As expected, quantization leads to the worst performance. 

\begin{figure}[htb] 

	\centering
	\begin{subfigure}[b]{0.44\textwidth}
		\includegraphics[width=\textwidth]{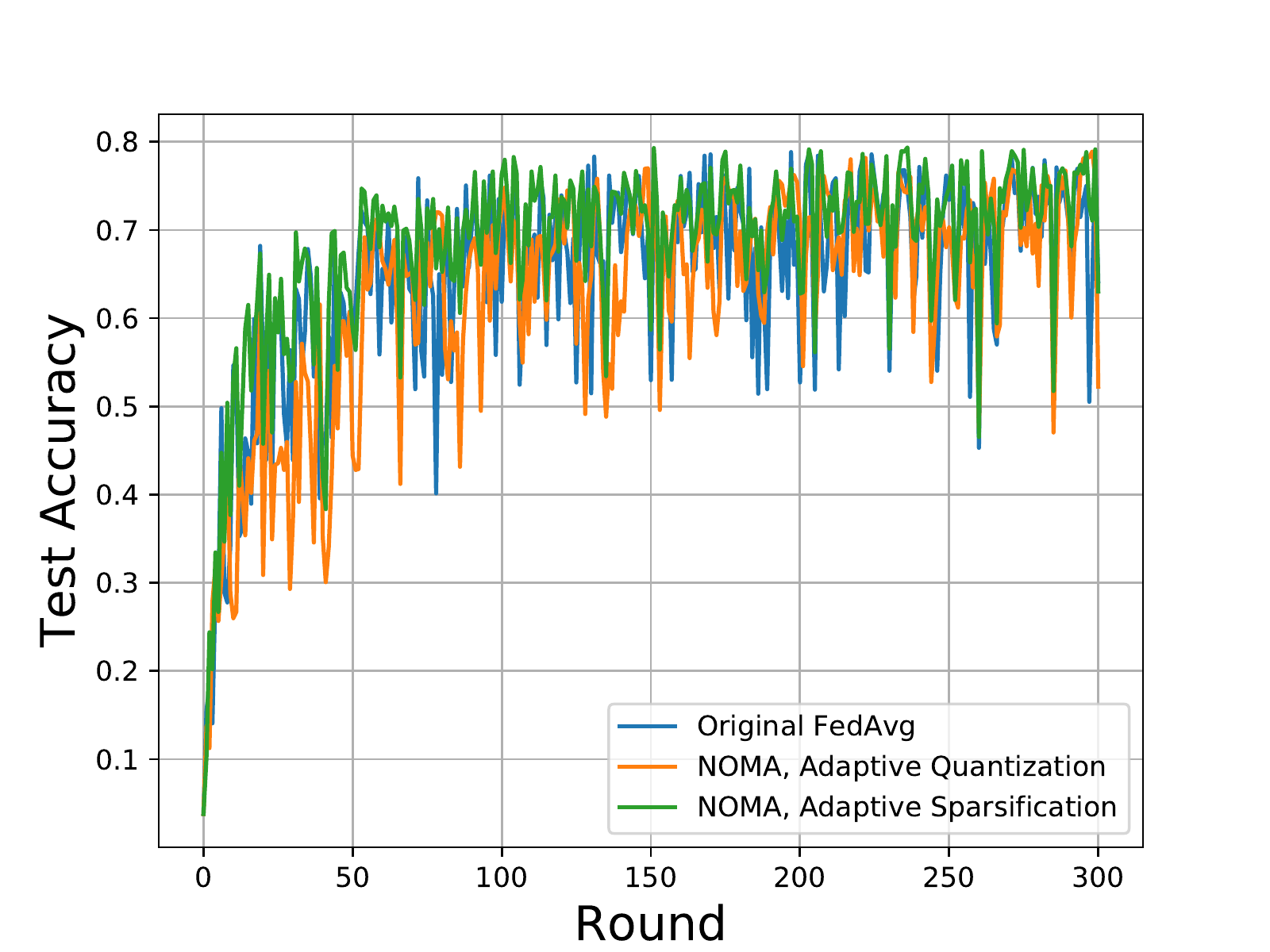}
	\end{subfigure}
	\begin{subfigure}[b]{0.44\textwidth}
		\includegraphics[width=\textwidth]{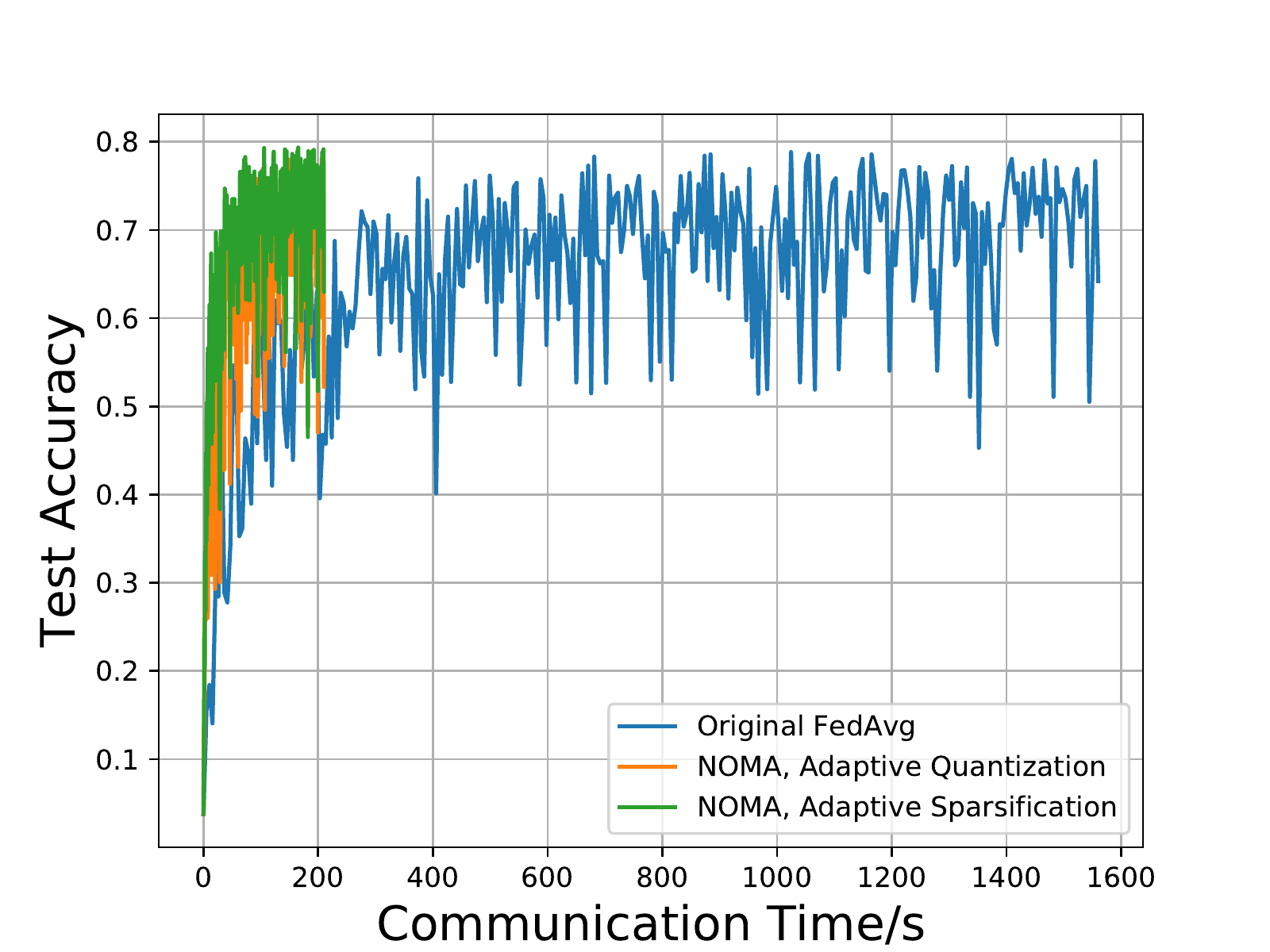}
	\end{subfigure}
	\caption{Test accuracy on  FEMNIST datasets. Left: Test accuracy comparison v.s. communication rounds. 
	Right: Test Accuracy of original TDMA-based FedAvg and NOMA compression based FedAvg update v.s communication time.}
\end{figure}

\begin{figure}[htb]
\centering
	\begin{subfigure}[b]{0.44\textwidth}
		\includegraphics[width=\textwidth]{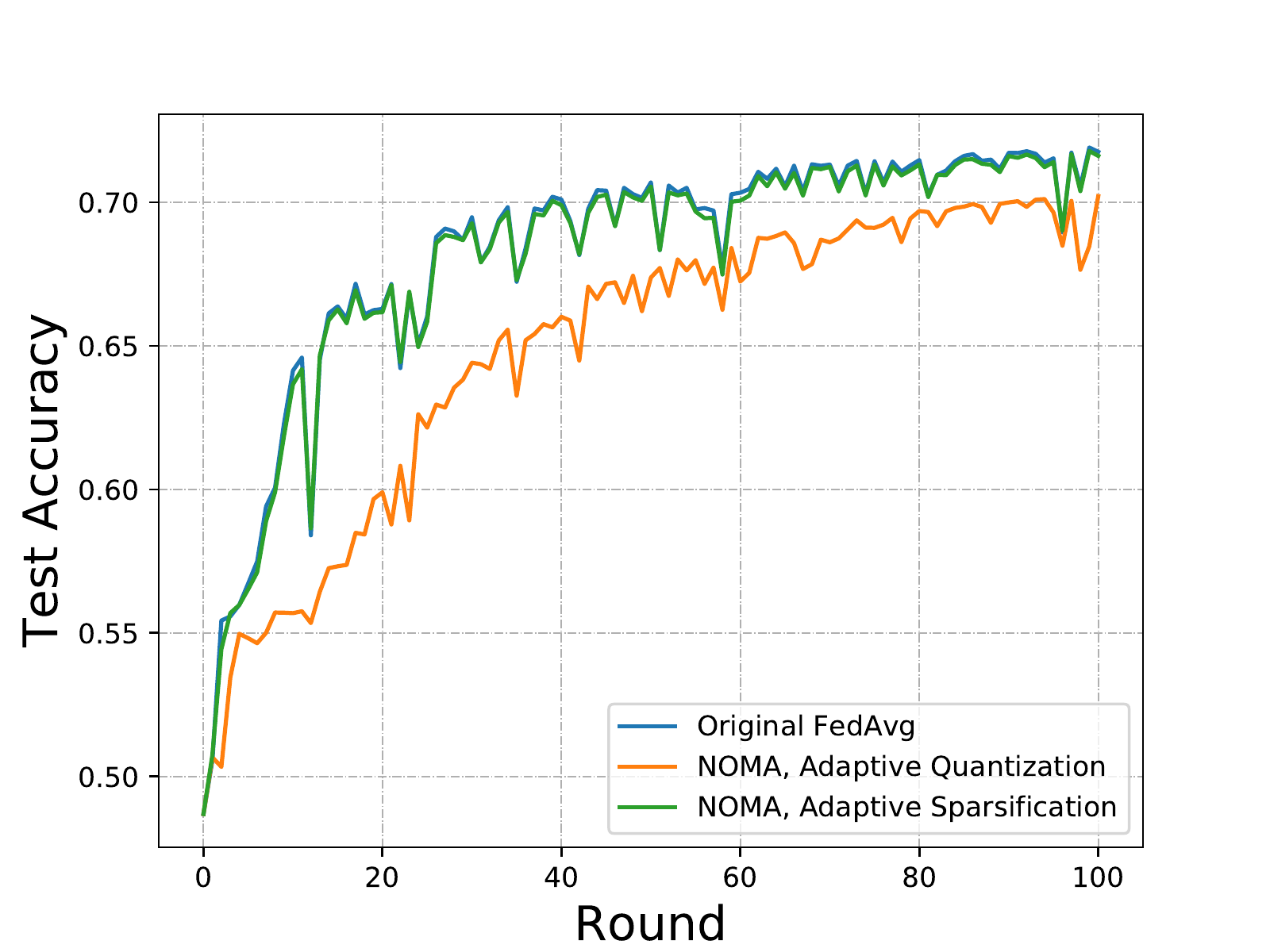}
		\label{fig:sent140_round}
	\end{subfigure}
	\begin{subfigure}[b]{0.44\textwidth}
		\includegraphics[width=\textwidth]{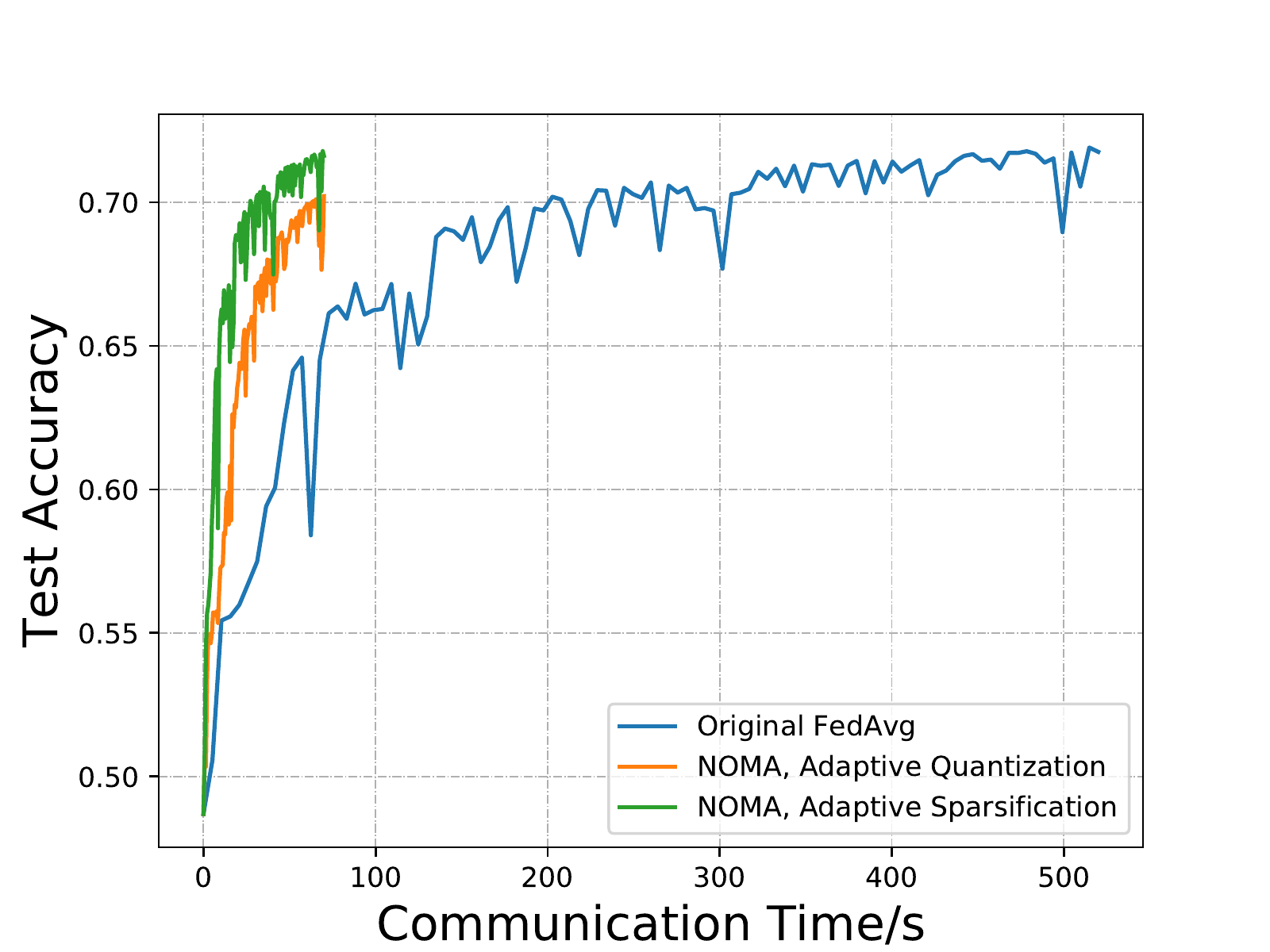}
		\label{fig:sent140_time}
	\end{subfigure}
	\caption{Test accuracy on  \emph{Sent140} datasets. Left: Test accuracy comparison v.s. communication rounds. 
	Right: Test Accuracy of original TDMA-based FedAvg and NOMA compression based FedAvg update v.s communication time.}
	\label{fig:other_datasets}
\end{figure}

From communication perspective, in the final round for FEMNIST, to get $79.5$\% of accuracy, TDMA-based original FedAvg takes approximately $1600$s, while both schemes with compression under NOMA  take about $200$s. Similarly, \emph{Sent140} achieves $73.5$\% accuracy with $75$s in NOMA compression based FedAvg and over $510$s with TDMA-based FedAvg. Notice that in \emph{Sent140} case, adaptive sparsification and original FedAvg have almost identical performance in each round. 
These results again prove the remarkable advantage of our proposed  NOMA-aided compression based update strategy, especially in terms of update latency.

\section{Conclusions}
In this work, we propose to apply NOMA in the FL for uplink model update. We consider wireless fading channels during the update process and adaptively compress gradient values according to either sparsification or quantization. Experiment results from three common dataset have verified our scheme and shown that NOMA-based strategy can significantly reduce communication time while preserving accuracy. One possible future direction is to apply NOMA power control in the uplink or use multiple antennas to further improve spectrum efficiency.


\begin{thebibliography}{13}

\bibitem{cache}
L. T. Tan and R. Q. Hu, ``Mobility-Aware Edge Caching and Computing in Vehicle Networks: A Deep Reinforcement Learning'', in \emph{IEEE Trans.  Veh. Technol.}, vol. 67, no. 11, pp. 10190-10203, Nov. 2018.

\bibitem{book}
I. Goodfellow, Y. Bengio, and  A. Courville, ``Deep Learning'', \emph{MIT Press}, 2016. 

\bibitem{fedavg}
H. B. McManhan, E. Moore, D. Ramage, S. Hampson, and B. A. Arcas, ``Communication-efficient learning of deep networks from decentralized data'',  in \emph{Proc. 20th International Conference on Artificial Intelligence and Statistics,} Fort Lauderdale, Florida, 2017. 

\bibitem{SHan}
Y. Lin, S. Han, H. Mao, Y. Wang, and W. J. Dally, 
``Deep gradient compression: reducing the communication bandwidth for distributed training'', in \emph{ICLR}, 2018. 


\bibitem{tian}
T. Li, A. K. Sahu, M. Zaheer, M. Sanjabi, A. Talwalkar, and V. Smith, ``Federated optimization in heterogeneous networks'', [Online]: https://arxiv.org/abs/1812.06127,  2019.

\bibitem{comm}
M. M. Amiri and D. Gündüz, ``Machine learning at the wireless edge: distributed stochastic gradient descent over-the-air'', in \emph{IEEE International Symposium on Information Theory,} Paris, France, Jul. 2019. 

\bibitem{uplink}
Z. Zhang, H. Sun and R. Q. Hu, ``Downlink and Uplink Non-Orthogonal Multiple Access in a Dense Wireless Network," in \emph{IEEE J. Sel. Area Comm.}, vol. 35, no. 12, pp. 2771-2784, Dec. 2017

\bibitem{noma}
N. Zhang, J. Wang, G. Kang, and Y. Liu, ``Uplink nonorthogonal multiple access in 5G systems'', in \emph{IEEE Commun. Lett.}, vol. 20, no. 3, pp. 458-461, Mar. 2016. 

\bibitem{Zhou}
S. Zhou	, Y. Wu, Z. Ni, X. Zhou, H. Wen, and Y. Zou,``Dorefa-net: Training
low bitwidth convolutional neural networks with low bitwidth gradients,'' [Online]: https://arxiv.org/abs/1606.06160, 2018


\bibitem{fairness}
Mohanad M. Al-Wani, A. Sali, \emph{et al.}, ``On short term fairness and throughput of user clustering for downlink non-orthogonal multiple access system'', in \emph{Proc. IEEE 89th Veh. Techn. Conf.}, 2019. 


\end{thebibliography}
\end{document}